\def\bd{\begin{displaymath}}\def\ed{\end{displaymath}}
\def\be{\begin{equation}}\def\ee{\end{equation}}
\def\bea{\begin{eqnarray}}\def\eea{\end{eqnarray}}
\def\ba{\begin{array}}\def\ea{\end{array}}
\def\nn{\nonumber}\def\lb{\label}

\def\a{\alpha}\def\b{\beta}\def\d{\delta}
\def\f{\phi}\def\g{\gamma}\def\h{\theta}
\def\k{\kappa}\def\l{\lambda}\def\n{\nu}\def\o{\omega}\def
\p{\pi}\def\r{\rho}\def\t{\tau}

\def\D{\Delta}\def\H{\Theta}

\def\inf{\infty}\def\id{\equiv}\def\mo{{-1}}

\def\eom{equations of motion }
\def\coo{coordinates }
\def\pb{Poisson brackets }

\def\min{Minkowski }

\def\PL#1{Phys.\ Lett.\ {\bf#1}}
\def\PRL#1{Phys.\ Rev.\ Lett.\ {\bf#1}}
\def\PR#1{Phys.\ Rev.\ {\bf#1}}\def\CQG#1{Class.\ Quantum Grav.\ {\bf#1}}
\def\NP#1{Nucl.\ Phys.\ {\bf#1}}

\def\JoP#1{J.\ Phys.\ {\bf#1}} \def\IJMP#1{Int.\ J. Mod.\ Phys.\ {\bf #1}}
\def\MPL#1{Mod.\ Phys.\ Lett.\ {\bf #1}} 
\def\PRep#1{Phys.\ Rep.\ {\bf#1}}
\def\AoP#1{Ann.\ Phys.\ {\bf#1}}
\def\JHEP#1{JHEP\ {\bf#1}}
\def\RMP#1{Rev.\ Mod.\ Phys.\ {\bf#1}}\def\AdP#1{Annalen Phys.\ {\bf#1}}
\def\arx#1{{\tt arXiv:#1}}

\documentclass[12pt]{article}

\def\eol{\sqrt{{E'}^2-\o_0^2l^2}\,}\def\ij{{ij}}\def\cE{{\cal E}}\def\ah{{\a^2\over2}\,}

\begin{document}
\begin{titlepage}
\title{Classical dynamics on curved Snyder space}

\author{ B. Iveti\'c\footnote{e-mail: bivetic@yahoo.com}, S. Meljanac\footnote{e-mail: 
meljanac@irb.hr}, \\ \small Rudjer Bo\v skovi\'c Institute, Bijeni\v cka c.\ 54, 
HR-10002 Zagreb, Croatia\and S. Mignemi\footnote{e-mail: smignemi@unica.it},
\\ \small Dipartimento di Matematica e Informatica,\\ \small
Universit\`a di Cagliari, viale Merello 92,\ I-09123 Cagliari, Italy\\
\small and INFN, Sezione di Cagliari}
\maketitle

\begin{abstract}
We study the classical dynamics of a particle in nonrelativistic
Snyder-de Sitter space. We show that for spherically symmetric systems, parametrizing
the solutions in terms of an auxiliary time variable, which is a function only of the
physical time and of the energy and angular momentum of the particles, one can reduce
the problem to the equivalent one in classical mechanics.
We also discuss a relativistic extension of these results, and a generalization to the
case in which the algebra is realized in flat space.
\end{abstract}

\end{titlepage}
\section{Introduction}
In recent times, the hypothesis that spacetime has a noncommutative structure
is becoming rather popular and has been investigated from several perspectives,
leading to a variety of models \cite{ncg}.
Among these proposals, the Snyder model \cite{Sn} has the remarkable property of
leaving the Lorentz invariance unaltered.
In its classical limit the noncommutativity emerges in dimensions $N\ge2$
from the nontrivial \pb of the position coordinates, that depend on a scale $\b$,
usually identified with the inverse of the Planck  mass.
Because of the presence of this fundamental scale, the Snyder model can be interpreted
as an example of doubly special relativity \cite{dsr}.
As such, it can be realized in a curved momentum space \cite{KG}.
Special relativity is recovered in the limit $\b\to 0$.
Several investigations on its properties have been carried out in the last years
\cite{BK,Mi,Mi1,snyder,SM1},
and also a generalization interpolating with $\k$-\min space has been proposed \cite{MS}.

More recently, the proposal was advanced to extend the Snyder model to spacetimes of
constant curvature, by introducing a new fundamental constant $\a$, whose square is
proportional to the inverse of the cosmological constant \cite{KS}. This extension
was motivated by arguments coming from loop quantum gravity \cite{KS}. The model
(originally named triply special relativity because it contains the three fundamental
scales $c$, $\a$ and $\b$) presents an interesting
duality between position and momentum space, both being realized as spaces of constant
curvature \cite{SM}. For this reason we shall refer to it as Snyder-de Sitter (SdS)
model, although in this paper we study its Euclidean version. Its properties have been
investigated in a number of papers \cite{SM,tsr}.

In particular, the one-particle nonrelativistic classical and quantum dynamics were
studied in \cite{SM}. In that paper, the investigation of the classical dynamics
was limited to one dimension. Explicit solutions were found for the free particle and
the harmonic potential, and the qualitative difference between models with opposite
sign of the coupling constants was highlighted.

In the present paper, we extend the discussion of the classical dynamics to generic
dimensions. This generalization is especially interesting because
the noncommutativity of the coordinates is not effective in one dimension.
From our investigation will result however that the radial equation of motion in
higher dimensions does not substantially differ from the one-dimensional equation.
Moreover, for spherically symmetric potentials it is always possible
to find the solutions in terms of those of classical
mechanics, using an energy-dependent rescaling of the time parameter.
Remarkable differences arise instead in the angular sector. In this case improving
terms must be added to the potential if one wants to extend the above property
to the angular variables.

We also  give a relativistic extension of the dynamics, in the hypothesis that the
time components of the position and momentum vectors obey canonical \pb with the
other variables.

The plan of the paper is the following: in sect.\ 2 we review the classical
version of the nonrelativistic SdS model. In sect.\ 3 we recall the geometry of
spaces of constant curvature, introducing projective coordinates. In sect.\ 4 we
present the Hamiltonian formalism for our model and in sect.\ 5 determine the
2-dimensional free particle motion in  polar coordinates. Sect.\ 6 is dedicated
to the 2-dimensional harmonic oscillator, and sect.\ 7 to the Kepler problem.
In sect.\ 8 we briefly discuss
some generalizations to higher  dimensions and to relativistic dynamics.
Finally, in sect.\ 9 we consider the case in which the SdS algebra is realized
in a flat background.

\section{The model}
Let us consider the $N$-dimensional nonrelativistic SdS model. This is defined on a
$2N$-dimensional phase space with position coordinates $x_i$ and momentum \coo
$p_i$ ($i=1,\dots,N$), with fundamental \pb
\bd
\{x_i,x_j\}=\b^2J_{ij},\qquad\qquad\{p_i,p_j\}=\a^2J_{ij},
\ed
\be\lb{algebra}
\{x_i,p_j\}=\d_{ij}+\a^2x_ix_j+\b^2p_ip_j+2\a\b\,p_ix_j,
\ee
where $J_{ij}=x_ip_j-x_jp_i$ are the generators of the group of rotations, and $\a$ and
$\b$ are two coupling constants with dimension of inverse length and inverse mass,
respectively\footnote{We adopt units in which $c=G=1$.}. Of course, these \pb obey the
Jacobi identity and therefore define a consistent symplectic structure on phase space.

According to these definitions, the $J_{ij}$ obey the usual \pb of the generators of the
$N$-dimensional rotation group, while $x_i$ and $p_i$ transform as $N$-vectors under
the same group. The rotational invariance is therefore preserved. However, the translation
invariance, generated by the $p_i$, is realized in a nonlinear way on position space
\cite{BK,Mi},
as is evident from (\ref{algebra}). In the limit $\a\to0$, the algebra (\ref{algebra})
reduces to that of the ordinary nonrelativistic Snyder model, while for $\b\to0$ it is
the symmetry algebra of a sphere endowed with projective coordinates.

An alternative possibility is that the sign in front of the coupling constants be reversed,
and therefore
\bea\lb{aalgebra}
\{x_i,x_j\}&=&-\b^2J_{ij},\qquad\qquad\{p_i,p_j\}=-\a^2J_{ij},\cr
\{x_i,p_j\}&=&\d_{ij}-\a^2x_ix_j-\b^2p_ip_j-2\a\b\,p_ix_j,
\eea
This can be seen as an analytical continuation of the previous model for $\a\to i\a$ and
$\b\to i\b$, and was called anti-Snyder-de Sitter (aSdS) model in \cite{SM}.
Notice that it is not possible to make the analytic continuation of only one
of the coupling constants. Moreover, in this case the bound $(\a x_k+\b p_k)^2<1$ must be
imposed in phase space \cite{SM}, in analogy with other models derived from doubly special
relativity. The limit $\a=0$ corresponds to the nonrelativistic
anti-Snyder model in flat space \cite{SM1}, while $\b=0$ is the symmetry algebra of an
hyperbolic space.

As is evident from the previous discussion,
the two models have rather different physical properties \cite{SM}.
Here, we discuss in detail the first case, and only give some hints on the second.

\section{Geometry}
In order to better understand the behavior of the trajectories in space, it is
useful to discuss the geometry of the model \cite{Mi1}. In particular, the algebra
generated by
the $J_{ij}$ and $p_i$ is $SO(N+1)$ for $\a^2>0$ or $SO(N,1)$ for $\a^2<0$ and
describes the spatial symmetries of an $N$-sphere or pseudosphere, respectively.

The metric of the $N$-sphere can be obtained by embedding it in an $(N+1)$-dimensional
Euclidean space,
with coordinates $X_\a$, with $\a=1,\dots,N+1$, on which is imposed the constraint
$X_\a^2=1/\a^2$.
If one chooses projective coordinates $x_i=X_i/(\a X_{N+1})$ to parametrize the
sphere, its metric takes the form
\be\lb{metric}
g_{ij}={(1+\a^2x^2)\d_{ij}-\a^2x_ix_j\over(1+\a^2x^2)^2},
\ee
with inverse
\be
g^\ij=(1+\a^2x^2)(\d^{ij}+\a^2x^ix^j).
\ee
As is well known, a single parametrization cannot be regular on the whole sphere.
In particular, in these \coo the north and south poles $X_{N+1}=0$ correspond to
$x_i\to\inf$.

It is easy to see that the symmetry group of this metric is $SO(N+1)$, generated by the
rotation generators $J_{ij}$, and the translation generators $p_i=J_{i,N+1}$ satisfying
the algebra (\ref{algebra}) with $\b=0$. The algebra admits the Casimir operator
$p^2+\ah J^2$, with $J^2=J_\ij J^\ij$.
\medskip

When $\a^2<0$, the relevant manifold is an $N$-dimensional pseudosphere, that can be
constructed as an hyperboloid of equation $X_i^2-X_{N+1}^2=-1/\a^2$
in an $(N+1)$-dimensional Minkowski space, with timelike coordinate $X_{N+1}$.
Choosing projective coordinates as before, one obtains
the metric
\be
g_{ij}={(1-\a^2x^2)\d_{ij}+\a^2x_ix_j\over(1-\a^2x^2)^2},
\ee
which is the analytic continuation of (\ref{metric}) for $\a\to i\a$, and whose inverse
is
\be
g^\ij=(1-\a^2x^2)(\d^{ij}-\a^2x^ix^j).
\ee
Now one must impose $x^2\le1/\a^2$, and the points for which $x^2=1/\a^2$ correspond
to infinity of the hyperboloid.

In this case the symmetry group of the metric is $SO(N,1)$, generated by the
rotation generators $J_{ij}$ and the translation generators $p_i$, that satisfy the
algebra (2) with $\b=0$. Also this algebra admits a Casimir operator $p^2-\ah J^2$.

It is  interesting to observe that an analogous construction can be done for the
momentum space \cite{KG}. In that case the symmetries are generated by the $J_\ij$
and the positions $x_i$, and satisfy the algebra (\ref{algebra}) or (\ref{aalgebra})
with $\a=0$. Combining with spatial symmetries, it is easy to see that the phase
space is invariant under $SO(N+2)$ for SdS, or $SO(N+1,1)$ for aSdS. Also these
symmetries are realized nonlinearly.

\section{Dynamics}
In \cite{SM} the classical dynamics of the nonrelativistic SdS model was investigated
in one dimension. The solutions for a free particle and a harmonic oscillator were found.
Here we generalize these solutions to $N$ dimensions, and show that for spherically
symmetric potentials the problem can be reduced to the study of a radial function,
whose equation of motion can be reduced to the classical one by a suitable
coordinate-dependent reparametrization of time.

Due to the nontrivial symplectic structure, it is convenient to write down the dynamics
in Hamiltonian form. The first step is the choice of a Hamiltonian. This choice
is an additional input to the theory, independent of the symplectic structure, and
must be consistent with the identification of $x$ and $p$ as the physical
position and momentum. However, this requirement does not fix the form of the
Hamiltonian uniquely.

For the free particle, the most trivial choice would be
\be\lb{flatham}
H={p^2\over2m}.
\ee
However, in a curved space it is more appropriate to define the Hamiltonian as
\be
H={1\over2m}g^{ij}p_ip_j.
\ee
This is also equivalent to choosing the Casimir operator of the spacetime symmetry group
$SO(N+1)$ as Hamiltonian. In particular, in our case one has
\be\lb{ham}
H={1\over2m}\left[\left(1+\a^2x^2\right)p^2-\a^2(x\cdot p)^2\right]
={1\over2m}\left(p^2+\ah J^2\right).
\ee

In principle one may add to the Hamiltonian further correction terms proportional to
$\b$, in order to improve its invariance under the symmetries of phase space.
For example, in \cite{Mi,Mi1} considerations based on the momentum space invariance
suggested the
definition of a Hamiltonian $H=p^2/(1+\b^2p^2)$ for the Snyder model in flat space, with
straightforward generalization to curved space \cite{Mi1}.
Since this choice does not affect the main results of this paper we will use the simpler
forms (\ref{flatham}) or (\ref{ham}) adopted by most authors
\cite{KS,tsr,snyder}, which is closer in spirit to Snyder's original idea of deforming
only the Heisenberg commutation relations. An analogous choice has also been adopted for
more general models of nonrelativistic dynamics on noncommutative space \cite{das}.

\section{Free particle dynamics in two dimensions}
In two dimensions, the free particle dynamics can be investigated adopting polar coordinates.
One defines
\bd
\r=\sqrt{x_1^2+x_2^2},\qquad\h=\arctan{x_2\over x_1},
\ed
\be\lb{polar}
p_\r={x_1p_1+x_2p_2\over\sqrt{x_1^2+x_2^2}},\qquad p_\h\id J_{12}=x_1p_2-x_2p_1.
\ee
The nonvanishing \pb for these \coo read (notice that the transformations (\ref{polar}) are not
symplectic)
\bd
\{\r,\h\}=\b^2{p_\h\over\r},\qquad\{\h,p_\r\}=\b^2{p_\r p_\h\over\r^2}+2\a\b{p_\h\over\r},
\ed
\be\lb{pb}
\{\r,p_\r\}=1+\a^2\r^2+\b^2\left(p_\r^2+{p_\h^2\over\r^2}\right)+2\a\b\r p_\r,\qquad\{\h,p_\h\}=1.
\ee
Since $\r$ and $p_\r$ transform as scalars, their \pb with $p_\h$ vanish.

In these coordinates, the Hamiltonian (\ref{ham}) reads
\be
H={1\over2m}\left(p_\r^2+{p_\h^2\over\r^2}+\a^2p_\h^2\right).
\ee
Using (\ref{pb}) one can then write down the Hamilton equations, that take a very simple form,
namely
\bd
\dot\r={\D\over m}\,p_\r,\qquad\dot p_\r={\D\over m}\,{p_\h^2\over\r^3},
\ed
\be
\dot\h={\D\over m}\,{p_\h\over\r^2},\qquad\dot p_\h=0,
\ee
with
\be\lb{delta}
\D=1+\a^2\r^2+\b^2\left(p_\r^2+{p_\h^2\over\r^2}\right)+2\a\b\r p_\r.
\ee
The form of these equations is exactly the same as in the classical case, except for the
factor $\D$.
One can however get rid of it by defining an auxiliary time variable $\t$ such that
\be\lb{tau}
d\t=\D dt.
\ee
In terms of the variable $\t$ the equations are the same as in classical mechanics,
and hence have the same solution. In particular, the angular momentum $p_\h$ is conserved.

The solution of the
SdS model can therefore simply be obtained by integrating (\ref{tau}) from the classical
solution, which can in turn be derived from the conservation of the Hamiltonian.
Calling $E$ the value of the Hamiltonian and $l$ that of the angular momentum $p_\h$,
one has
\be
2E={1\over m}\left(p_\r^2+{l^2\over\r^2}+\a^2l^2\right).
\ee
Substituting $p_\r=m\,d\r/d\t$,
\be
{d\r\over d\t}={1\over m}\sqrt{2mE'-{l^2\over\r^2}},
\ee
where $E'=E-\a^2l^2/2m$. Integrating, one obtains\footnote{All solutions we discuss admit
an arbitrary choice of the origin of time. For simplicity we shall always set to zero this
integration constant.}
\be\lb{fresol}
\r^2={1\over m}\left(2E'\t^2+{l^2\over2E'}\right),\qquad
p_\r^2={8m{E'}^3\t^2\over4{E'}^2\t^2+l^2}.
\ee
 Hence
\be
\D=1+2\b^2mE'+{\a^2\over m}\left(2E'\t^2+{l^2\over2E'}\right)+4\a\b E'\t,
\ee
and then
\be
t=\int{d\t\over\D}={1\over\l}\ \arctan\left[{2\a E'\over\l}\left({\a\over m}\,\t+\b\right)\right],
\ee
where
\be
\l=\a\sqrt{{2\over m}\left(E'+{\a^2l^2\over2m}\right)}=\a\sqrt{2E\over m}.
\ee
Inverting,
\be\lb{tfree}
\t={m\over\a}\left({\l\over2\a E'}\tan\l t-\b\right),
\ee
 and substituting back in (\ref{fresol}) one has
\be
\r^2={1\over\a^2}\left(\sqrt{E\over E'}\ \tan\l t-\b\sqrt{2mE'}\right)^2+{l^2\over2mE'}.
\ee
The radial coordinate $\r$ diverges when the argument of the tangent is $\pm{\p\over2}$.
However, these points are simply the poles of the 2-sphere. In fact, the proper distance
along a radial trajectory is given by
\be
s=\int ds=\int{d\r\over1+\a^2\r^2}={\arctan\a\r\over\a},
\ee
which is always finite for finite $t$. For $l=0$, the solution has the same form as
in one dimension \cite{SM}.

The solution for the angle $\h$ can be obtained in a analogous way. One has
\be
\h=\int {l\over\r^2}\,d\t=\arctan{2E'\t\over l},
\ee
with $\t$ given by (\ref{tfree}).
\medskip

A similar calculation can be done in the aSdS case. The results can be more easily obtained
by analytic continuation of the previous ones.
In particular, the solution (\ref{fresol}) is still valid, but now $E'=E+\a^2l^2$ and
\be
\t={m\over\a}\left({\l\over2\a E'}\tanh\l t-\b\right),
\ee
It is easy to check that the bound $(\a x_k+\b p_k)^2<1$ is satisfied.

\section{Harmonic oscillator in two dimensions}
We consider now the simplest example of an interacting particle, given by the harmonic
oscillator. Starting from the Hamiltonian
\be
H={1\over2m}\left(p^2+\ah J^2\right)+{m\o_0^2\over2}x^2,
\ee
and going to polar coordinates, one has
\be
H={1\over2m}\left(p_\r^2+{p_\h^2\over\r^2}+\a^2p_\h^2\right)+{m\o_0^2\over2}\r^2,
\ee
and hence
\bd
\dot\r={\D\over m}\, p_\r,\qquad\dot p_\r={\D\over m}\left({l^2\over\r^3}+\o_0^2\r\right),
\ed
\be
\dot\h={\D\over m}\,{p_\h\over\r^2}-\b^2m\o_0^2p_\h,\qquad\dot p_\h=0,
\ee
where  $\D$ is given by (\ref{delta}).
The form of the radial equations is
still the same as in the classical case, except for the factor $\D$, while the angular equation
has a more general form. In the following we shall show how to modify the Hamiltonian in order
to have an analogous equation also for the angular variable.
For the moment, we proceed to solve the radial equation by the substitution $dt=\D d\t$.

The classical solution can be obtained from the conservation of the Hamiltonian.
With the same notations as in the previous section,
\be
2E={1\over m}\left(p_\r^2+{l^2\over\r^2}+\a^2l^2\right)+m\o_0^2\r^2,
\ee
and therefore
\be
{d\r\over d\t}={1\over m}\sqrt{2mE'-{l^2\over\r^2}-m^2\o_0^2\r^2}.
\ee
Integrating, one gets
\be\lb{oscsol}
\r^2={1\over m\o_0^2}\left[E'-\eol\cos2\o_0\t\right],
\qquad p_\r^2={{E'}^2-\o_0^2l^2\over\o_0^2\r^2}\sin^22\o_0\t,
\ee
and hence
\bea
\D&=&1+\left(m\b^2+{\a^2\over m\o_0^2}\right)E'+\left(m\b^2-{\a^2\over m\o_0^2}\right)
\eol\cos2\o_0\t\cr&&+{2\a\b\over\o_0}\eol\sin2\o_0\t.
\eea

From (\ref{tau}), after some straightforward algebraic manipulations, one obtains
\be\lb{tosc}
\tan\o_0\t={\k\tan\o t-{2\a\b\over\o_0}\eol\over1+\left(\b^2m+{\a^2\over m\o_0^2}\right)E'-
\left(\b^2m-{\a^2\over m\o_0^2}\right)\eol}
\ee
where
\be\lb{kappa}
\k=\sqrt{1+2E'\left(\b^2m+{\a^2\over m\o_0^2}\right)+\o_0^2l^2
\left(\b^2m+{\a^2\over m\o_0^2}\right)^2},\qquad\o=\k\o_0
\ee
After substituting in (\ref{oscsol}) one trivially gets the solution. Since it is
very cumbersome, we do not report its explicit form here.
When $l=0$, (\ref{oscsol}) reproduces the one-dimensional solution \cite{SM}.
The solution is periodic in time, with frequency $\o$ that depends on the energy and increases
with it.

The equation for $\h$ is rather complicated. However, it can be simplified if one adopts a modified
Hamiltonian, as suggested in \cite{SM} for the quantum problem. This is obtained by adding a term
${\b^2\over4}m\o_0^2J^2$ to the original Hamiltonian, getting
\be\lb{oscham}
H={1\over2m}\left(p^2+\ah J^2\right)+{m\o_0^2\over2}\left(x^2+{\b^2\over2}\,J^2\right).
\ee
The new term renders the Hamiltonian invariant under the full $SO(3)$ symmetry of
momentum space and therefore enhances its symmetry properties. In particular, the potential
has now vanishing \pb with the $x_i$. In the quantum case, the addition of this term permits to
preserve the degeneracy of the energy with respect  to angular momentum \cite{SM}.
The additional term only affects the Hamilton equation for the angle $\h$, that now
takes the simple form
\be\lb{angeq}
\dot\h={\D\over m}\,{p_\h\over\r^2}
\ee

The solutions can then be obtained as above introducing the auxiliary variable $\t$.
 The solution (\ref{oscsol}), (\ref{tosc})
of the radial equation is unchanged, except for the substitution $E'\to E''=E'-\b^2m\o_0^2l^2/2$.
The solution of (\ref{angeq}) is instead
\be
\h=\arctan\left[{E''+\sqrt{{E''}^2-\o_0^2l^2}\over\o_0l}\,\tan\o_0\t\right],
\ee
with $\tan\o_0\t$ still given by (\ref{tosc}).

We also notice that if one writes down the equation for the trajectories ensuing from the
Hamiltonian
(\ref{oscham}), $d\r/d\h=\dot\r/\dot\h$, the $\D$ factors cancel out and one recovers the
classical equation. Hence the trajectories are still ellipses, although they are covered
with a velocity different from the classical one.

\medskip
In the aSdS case, the solution can simply be obtained by analytic continuation of (\ref{tosc})
and (\ref{kappa}) to negative values of $\a^2$ and $\b^2$.
The frequency of oscillation now decreases when the energy increases. It can be shown that
the bound $(\a x_k+\b p_k)^2<1$ is satisfied.

\section{The Kepler problem}
Several generalizations of the previous results are possible.
Of course, using polar \coo is especially useful for the study of spherically symmetric
potentials. In particular, given the Hamiltonian
\be
H={1\over2m}\left(p_\r^2+{p_\h^2\over\r^2}+\a^2p_\h^2\right)+V(\r),
\ee
the relation $\dot\r=\D p_\r/m$ still holds, and one can define as before an auxiliary
variable $\t$ in terms of which the system obeys the classical radial equation of motion.
Once the classical solution is known, one can obtain $\t(t)$
and hence the solution of the radial SdS equation, by integrating (\ref{tau}).

The solution of the angular equations instead is in general difficult, unless one
improves the symmetry of the potential by the substitution
$V(\r)\to V\left(\sqrt{\r^2+\b^2J^2/2}\right)$.
An example is given by the Kepler problem. For a potential $k/\sqrt{\r^2+\b^2J^2/2}$,
the calculations can be done as explained above, and in particular
the orbits maintain their classical shape, like for the harmonic oscillator. Only the
period of the orbits is changed and can be evaluated performing the integral of
$\D^\mo d\t$ over an orbit.

\medskip
Of course solutions can be found also if one chooses the traditional potential $V=-k/\r$,
but are much more involved. Let us consider the equation of the orbits in this case.
Since the angular momentum is conserved, the problem can be reduced to the motion on a plane.
In polar coordinates, the \eom are
\bd
\dot\r={\D\over m}\, p_\r,\qquad\dot p_\r={\D\over m}\left({p_\h^2\over\r^3}-{k\over\r^2}\right),
\ed
\be
\dot\h={\D\over m}\,{p_\h\over\r^2}-\b^2{kp_\h\over\r^3},\qquad\dot p_\h=0,
\ee
where  $\D$ is given as usual by (\ref{delta}). It is apparent that the angular momentum
component $p_\h=l$ is conserved.
Moreover, from the conservation of the Hamiltonian follows that
\be
\dot\r^2={\D^2\over m^2}\left(2mE'+{2km\over\r}-{l^2\over\r^2}\right),
\ee
where $E'=E-\a^2l^2/2m$. Since at leading orders the parameter $\a$ is not relevant in the
following calculations, we set it to 0. Then,
\be
\left({1\over\r^2}{d\r\over d\h}\right)^2={2mE+2km/\r-l^2/\r^2\over l^2(1-\b^2km/\D\r)^2}.
\ee
Defining $u=1/\r$ and expanding at first order in $\b^2$,
\be\lb{kep1}
\left({du\over d\h}\right)^2\approx\left({2mE'\over l^2}+{2km\over l^2}u-u^2\right)(1+2\b^2kmu).
\ee
Deriving with respect to $\h$, eq.\ (\ref{kep1}) becomes
\be\lb{kep2}
{d^2u\over d\h^2}={km\over l^2}\left(1+2\b^2mE'\right)-\left(1-4\b^2{k^2m^2\over l^2}\right)u
-3\b^2kmu^2,
\ee
and can be solved perturbatively in $\b^2$, expanding $u$ as $u=u_0+\b^2u_1+\dots$.
At order 0, the equation admits the well-known solution
\be
u_0={mk\over l^2}(1+e_0\cos\h),
\ee
where $e_0=\sqrt{1+2El^2/mk^2}$ is the eccentricity of the orbit.
Substituting in (\ref{kep2}), one obtains for the first-order correction
\be
u_1=-{m^2k\over l^2}\left[E+{k^2m\over2l^2}+{k^2me_0\over l^2}\,\h\sin\h-
{k^2me_0^2\over2l^2}\cos2\h\right],
\ee
and hence
\be
u\approx{mk\over l^2}\left[1-\b^2m^2\left({E\over m}+{k^2\over2l^2}\right)
+e_0\cos\left(1+\b^2m^2{k^2\over l^2}\right)\h+\b^2m^2{k^2e_0^2\over2l^2}\cos2\h\right].
\ee
It follows that a perihelion shift $\d\h=-2\p\b^2m^2k^2/l^2$ arises. A similar result was
obtained in \cite{lei} using a different approach.

This result may be applied to planetary orbits. In this case, $k=GmM$, with $G$ the Newton constant,
$M$ the mass of the sun and $m$ that of the planet. Comparing with the known discrepancy from
the prediction of general relativity of the perihelion precession of Mercury, of order $10^{-12}$
rad/rev \cite{rom}, an upper limit of order $10^{-52}$ s/kg can be obtained for $\b^2$.
This should be compared with the value expected by
dimensional arguments, $\b^2\sim\hbar/c^2M_{Pl}^2\sim10^{-34}$s/kg, where $M_{Pl}$ is the Planck mass.
Thus, if one assumes that Snyder mechanics holds at planetary scales, the parameter $\b$ must be
nearly nine orders of magnitude less than its natural scale. Equivalently, the length scale of the
noncommutativity should be less than $10^{-9}$ Planck lengths.

It may be interesting to compare these results with the ones obtained in the case of a
Lie-algebra noncommutative model \cite{rom}. In that case, the noncommutativity can be
parametrized by a constant $\H$, which, due to the structure of the algebra, has different
dimension from $\b^2$. In fact, it turns out  that $\d\h=2\p\H mk/l^3$ and
$\H\sim10^{-64}$m$^2\sim10^7L_{Pl}^2$ \cite{rom}. In this case, the astronomical constraints on
the value of the noncommutative parameter are much weaker, but the sign of the correction is
opposite to the observed one.

Before concluding that observations do not support Snyder mechanics, it must be observed however
that its range of validity is usually assumed to include only systems close to the Planck scale,
otherwise paradoxical effect, like the "soccer ball" problem of doubly special relativity
\cite{sbp}, may arise.

\section{Generalizations}
The methods of the previous section can also be generalized to higher dimensions.
The equations of motion still coincide with the classical ones if
written in terms of the auxiliary variable $\t$. For example, in cartesian \coo
the free particle Hamiltonian (\ref{ham}) yields the equations
\be
\dot x_i={\D\over m}\,p_i,\qquad\dot p_i=0,
\ee
while for the harmonic oscillator with Hamiltonian (\ref{oscham}),
\be
\dot x_i={\D\over m}\,p_i,\qquad\dot p_i=-m\o_0^2\D\,x_i,
\ee
Analogously, in polar coordinates the radial equation does not depend on the dimension
$N$, while the angular equations have obvious generalizations, as in classical mechanics.
\medskip

One can also define a relativistic extension of the model. This is obtained by assuming
that the timelike \coo $x_0$ and $p_0$ satisfy canonical \pb with all the variables.
Of course this assumption is in contrast with the postulates of the original Snyder
model, in which time has nonvanishing \pb with the spatial variables,
and whose dynamics will be studied elsewhere \cite{Mi2}.
Like in special relativity, the Hamiltonian for a free particle in
laboratory \coo is given by its relativistic energy $p_0$, or using the mass shell
constraint,
\be\lb{relham}
H=p_0=\sqrt{p_i^2+\ah J_\ij^2+m^2}.
\ee
In particular, if one defines an effective mass $m^\ast=\sqrt{m^2+\a^2J^2/2}$,
adapted to curved spaces (i.e. derived from the Casimir operator of the spacetime
symmetry group), the Hamiltonian takes its usual relativistic form.
Due to the conservation of energy, $H$ equals a constant $\cE$.

From (\ref{relham}) follow the \eom
\be
\dot x_i={\D\,p_i\over\cE},\qquad\dot p_i=0,
\ee
where $\D$ is defined as before.
In polar \coo the equations for the position variables read
\be
\dot\r={\D p_\r\over \cE},\qquad\dot\h={\D p_\h\over\cE\r^2}.
\ee
It follows that also in the relativistic case the equations can be solved using
the techniques developed above.

\section{Free particle in flat space}
The $\a$-modification of the \pb might have a different origin from the curvature of
spacetime, as in some noncommutative models coupled to magnetic fields \cite{DH}.
 It may therefore be interesting to consider the case in which the brackets
(\ref{algebra}) are realized in flat space.
In this case it is natural to adopt the Hamiltonian (\ref{flatham}) instead of
(\ref{ham}) for the free particle.
Due to (\ref{algebra}), this Hamiltonian has nonvanishing \pb with the momenta $p_i$,
which therefore are not conserved.

To solve this problem it turns out to be convenient to use cartesian coordinates.
This also allows one to work in generic dimension $N$.
From the rotational invariance follows that all $J_{ij}$ are conserved.
Moreover, conservation of energy implies
\be\lb{ener}
p^2=2mE.
\ee

The equations for the momenta derived from (\ref{flatham}) give
\be\lb{eqmom}
\dot p_i={\a^2\over2m}J_{ij}p_j.
\ee
Taking the derivative, one gets after some algebraic manipulations,
\be
\ddot p_i=-{\a^4J^2\over2m^2}p_i,
\ee
These equations can be immediately integrated, yielding
\be\lb{psol}
p_i=\sqrt{2mE}\left[A_i\sin(\n t+\f)-B_i\cos(\n t+\f)\right],
\ee
where $\n=\a^2J/\sqrt2m$, with $J=\sqrt{J_{ij}^2}$, and the constants $A_i$ and $B_i$
can be determined in terms of the  $J_{ij}$ from (\ref{eqmom}) and (\ref{ener}).
It follows that in the absence of the term $\a^2 J^2/2$ in the Hamiltonian, the
momenta rotate with frequency $\n$.

The equations for the positions read
\be\lb{eqpos}
m\dot x_i=\left(1+2\b^2mE+2\a\b D\right)p_i+\a^2Dx_i
\ee
where $D=x\cdot p\ $ and we have used (\ref{ener}).
From (\ref{eqmom}) and (\ref{eqpos}) follows that
\be
\dot D={\a^2\over m}D^2+4\a\b ED+2E(1+2\b^2mE)+{\a^2J^2\over2m}
\ee
This equation holds independently of the dimension of space.
The solution is
\be\lb{dsol}
D={m\g\over\a^2}\tan\g t-2mE{\b\over\a}
\ee
where $\g={\a\over m}\sqrt{2mE+\a^2J^2/2}$.
The $x_i$ can finally be determined algebraically substituting the solutions
(\ref{psol}) and (\ref{dsol}) into the definition of the $J_\ij$.

For example, in two dimensions one gets
 \bea
 &p_1=\sqrt{2mE}\sin(\n t+\f),\ &x_1={1\over\a}\left(\sqrt{\bar E\over E}\tan{\g t}-
\b\sqrt{2mE}\right)\sin(\n t+\f)+{l\over\sqrt{2mE}}\cos(\n t+\f),\cr
 &p_2=\sqrt{2mE}\cos(\n t+\f),\ &x_2={1\over\a}\left(\sqrt{\bar E\over E}\tan{\g t}-
\b\sqrt{2mE}\right)\cos(\n t+\f)-{l\over\sqrt{2mE}}\sin(\n t+\f),\nn
\eea
where $\bar E=E+\a^2l^2/2m$, $l^2=J^2/2$ and $\f$ is an integration constant.
(Note that the definition (\ref{ener}) of the energy $E$ is different from that of
sec.\ 5). This is a combination of a motion along a straight line with a rotation
around the origin. Some care must be taken however because $x_i$ diverges for a
finite value of $t$.

Of course, one can recover the results of sect.\ 5 using cartesian \coo as above.
In that case the momenta are conserved, and it is easy to see that the solution
is obtained from the previous one taking the limit  $\n\to0$.

\section{Conclusions}
We have calculated exact solutions of the radial equation of motion of the
nonrelativistic SdS model in generic dimensions.
The results have been obtained explicitly for the free particle and the
isotropic harmonic oscillator, but can easily be extended to
any spherically symmetric potential.

Moreover, we have considered the possibility that the noncanonical \pb of the SdS
model be realized in flat space, as may happen in the presence of a magnetic
field \cite{DH}. In this case the momenta are not conserved
even for a free particle, but vary in a periodic way.

The main result of our investigation is that the radial motion of a particle
in a spherically symmetric potential in SdS space obeys the same equations of
classical mechanics when written in terms
of an auxiliary time variable $\t$, which is a function of the real time and
depends on the energy and the angular momentum of the particle. The same result
can be extended to the angular motion if one adds to the potential an improving
term proportional to the square of the angular momentum, which restores the
$SO(N+1)$ invariance of momentum space. In absence of the improving term,
it is still possible to obtain approximate solutions. As an example, we have
discussed the orbits of particles in a Newtonian potential.

One may interpret these results by saying
that time flows faster or slower depending on the value of the energy of the
particle, as in some doubly special relativity models \cite{dsr}.
Alternatively, one may invoke a sort of renormalization group flow, according to
which the effective coupling constants depend on the energy (cf eq.\ (\ref{kappa})).

From a formal point of view, it seems plausible that this property follows from
the existence of a Darboux transformation between the \coo $p_i$, $q_i$, $t$ and
canonical
\coo $P_i$, $Q_i$, $\t$ in a parametrized (i.e.\ with time taken as a canonical
variable) version of the model. However, we were not able to find its explicit
form.

All the results discussed are simply extended to the case of aSdS space,
which is obtained by analytic continuation of SdS. The dynamics is
similar, except that in aSdS space the positions and momenta \coo are constrained
to satisfy an upper bound \cite{SM}. This is especially evident for the free
motion.

To conclude, it must be noted that some arbitrariness is present in the
definition of the Hamiltonian for aSdS models in dimension higher than one.
The choice depends on the symmetries one wishes to preserve and different
Hamiltonians may lead to different physical results (see also \cite{SM}
for a discussion of the quantum problem). Of course, the correct choice
depends on the physical setting one is considering.

The results of this paper may also be extended to more general models, with
arbitrary symplectic structure obeying suitable physical assumptions. We plan
to discuss this topic in a future paper.

\section*{Acknowledgements}
SM wishes to thank Rudjer Bo\v skovi\'c Institute for kind hospitality during
the latest stages of this work. This work was supported by the Ministry of
Science, Education and Sports of the Republic of Croatia under contract
No.\ 098-0000000-2865.

\end{document}